\documentclass[showpacs,preprintnumbers,showkeys,superscriptaddress,amsmath,amssymb,nofootinbib]{revtex4-2}
\usepackage{color}
\usepackage{amsmath}
\usepackage{amssymb}
\usepackage[utf8]{inputenc}
\usepackage{amsfonts}
\usepackage{bm}
\usepackage{bbold}
\usepackage{dsfont}
\usepackage{pstricks,pst-coil}
\usepackage{graphics}
\usepackage{graphicx}
\usepackage{slashed}
\usepackage{enumitem}

\newcommand{\be}{\begin{equation}}
\newcommand{\ee}{\end{equation}}
\newcommand{\ba}{\begin{eqnarray}}
\newcommand{\ea}{\end{eqnarray}}

\def\ni{\noindent}

\begin{document}

%\begin{center}

%\title{A gravitation model in the Doplicher-Fredenhagen-Roberts noncommutative space-time }
%
\title{\Large Non-linear electrodynamics emerging from a Lorentz-symmetry violation scenario}

%\bigskip
%\bigskip

%\author{ J. M. A. Paix\~ao} \email{jeff@cbpf.br}
%\affiliation{Centro Brasileiro de Pesquisas F\'isicas, Rua Dr. Xavier Sigaud
%150, Urca, CEP 22290-180, Rio de Janeiro, Brazil}

\author{M. J. Neves} \email{mariojr@ufrrj.br}
%\affiliation{Department of Physics and Astronomy, University of Alabama, Tuscaloosa, Alabama 35487, USA}
\affiliation{Departamento de F\'{i}sica, Universidade Federal Rural do Rio de Janeiro, BR 465-07, 23890-971, Serop\'edica, RJ, Brasil}

\author{G. Peruzzo} \email{giovaniperuzzo@ufrrj.br}
%\affiliation{Department of Physics and Astronomy, University of Alabama, Tuscaloosa, Alabama 35487, USA}
\affiliation{Departamento de F\'{i}sica, Universidade Federal Rural do Rio de Janeiro, BR 465-07, 23890-971, Serop\'edica, RJ, Brasil}

%\author{Jorge B. de Oliveira}\email{jorgebernardo998@gmail.com}
%\affiliation{Departamento de F\'{i}sica, Universidade Federal Rural do Rio de Janeiro, BR 465-07, 23890-971, Serop\'edica, RJ, Brazil}

%\author{ L. P. R. Ospedal } \email{leoopr@cbpf.br}
%\affiliation{Centro Brasileiro de Pesquisas F\'isicas, Rua Dr. Xavier Sigaud
%150, Urca, Rio de Janeiro, Brazil, CEP 22290-180}

%\author{J. A. Helay\"el-Neto}\email{helayel@cbpf.br}
%\affiliation{Centro Brasileiro de Pesquisas F\'isicas, Rua Dr. Xavier Sigaud
%150, Urca, Rio de Janeiro, Brazil, CEP 22290-180}

%%%%%%%%%%%%%%%%%%%%%%%%%%%%%%%%%%%%%%%%%%%%%%%%%%%%%%%%%%%%%%%%%%%%%%%%%%%%%%%%%%%%%%%%%%%%%%%%%%%%%%%%%%%%%%%%%%%%%%%%%%%

\date{\today}

\begin{abstract}
\ni

The effective action at one-loop for the quantum electrodynamics (QED) in the presence of background 
vectors and tensors that break the Lorentz symmetry is calculated in this work. We highlight 
that just the Lorentz symmetry violation through the Dirac matrices in the fermionic sector 
is considered since investigations in others sectors of the extended QED has been explored in recent papers. 
Using the proper time method, we show that some background vectors and tensors do not contribute 
to the effective action at one-loop and in the linear approximation. Thereby, only two background tensors contribute to the 
effective action at one-loop for small components of these tensors, and that a new combination of cubic and quartic non-linear electrodynamics 
emerge in the weak field regime.

\end{abstract}

%\pacs{11.15.-q; 11.10.Ef; 11.10.Nx}


\maketitle

\pagestyle{myheadings}
\markright{NLED emerging from a LSV scenario}

%%%%%%%%%%%%%%%%%%%%%%%%%%%%%%%%%%%%%%%%%%%%%%%%%%%%%%%%%%%%%%%%%%%%%%%%%%%%%%%%%%%%%%%%%%%%%%%%%%%%%%%%%%%%%%%%%%%%%%%%%%%%%%%%%%%%%%%%%%%%%%%%%%%%%%%%%%%%%%%%%%%%%%%%%%%%%%%%%%%%%%%%%%%%%%
%\newpage

\section{Introduction}
Lorentz symmetry is one of the main basis for the understanding of the Standard Model (SM).
Around the scale of $246$ GeV, the SM describes with great precision the 
fundamental interactions in the nature. However, it is known today that the SM is an 
effective theory that works at low energy scale. Several extensions of the SM at 
high energy scale are candidates to new physics for the description of the fundamental interactions. 
One of them is associated with the Lorentz symmetry violation (LSV) in which it is introduced background 
vectors and tensors that break the Lorentz symmetry in all the sectors of the SM. 
This model is known as Standard Model Extension (SME) \cite{Colladay,Colladay2}. One of the motivations 
for the construction of the SME is associated with the limit of low energy in string theory that emerge 
naturally vectors and tensors that select preferred directions in the spacetime \cite{Samuel}. 
Noncommutative field theories \cite{Pospelov}, and quantum gravity also show scenarios where 
the LSV arises \cite{Moffat}. All the bounds on the components of the background vectors and tensors of the SME are 
collected in the ref. \cite{tables} through the diverse experimental tests as atomic spectroscopy, astrophysical observations, collider physics, and others. 
Decays of $Z$ and $W$ into leptons bounded the LSV vectors at $\lesssim 10^{-4} \, \mbox{GeV}^{-1}$ \cite{Yuri}, the electron-positron scattering into Higgs-$Z$ and $Z$-$Z$ 
bounded at $\lesssim 10^{-5} \, \mbox{GeV}^{-1}$ in the electroweak of the SME \cite{Philipe}. The electromagnetic sector of the SME also was well constrained due to precision 
of the experimental results in the quantum electrodynamics (QED). As example, the photon sector described by Chern-Simons-like term and CPT-odd Carroll-Field-Jackiw (CFJ) in $1+2$ 
dimensions is constrained by astrophysical observations \cite{Kostelecky2008,Kostelecky2007}. In $1+3$ dimensions, the photon sector in the presence of an CPT-odd CFJ term was constrained 
by energy spectrum of hydrogen and the generation of electric dipole momentum for charged leptons \cite{Yuri2016}.
In connection with the QED extended, the effective action for fermions in the presence of LSV parameters has been investigated in some sectors, see the ref. \cite{Araujo}. In this reference, the authors calculated 
the effective lagrangian considering just the LSV introduced in the fermion mass, namely, $m\,\mathds{1}+H_{\mu\nu}\,\sigma^{\mu\nu}/2$, for the $H_{\mu\nu}$-antisymmetric tensor that breaks the Lorentz symmetry. 
For a most general case in which more LSV parameters are added to mass fermion and that it depends on the space-time coordinates, the effective action was calculated in the ref. \cite{Wagno}. Motivated by all these 
references, we calculate the effective action at one-loop for the fermion sector of QED extended when the LSV parameters are introduced directly in the Dirac matrices, \emph{i.e.}, the Dirac matrices are modified by
$\Gamma^{\nu}=\gamma^{\nu}+c^{\mu\nu}\gamma_{\mu}+d^{\mu\nu}\,\gamma_{\mu}\,\gamma_{5}+e^{\nu}\,\mathds{1}
+i\,f^{\nu}\,\gamma_{5}+\frac{1}{2} \, g^{\lambda\mu\nu}\,\sigma_{\lambda\mu}$, where $\left\{ \, c^{\mu\nu} \, , \, d^{\mu\nu} \, , \, e^{\nu} \, , \, f^{\nu} \, , \, g^{\lambda\mu\nu} \, \right\}$ 
have dimensionless components, and we keep the fermion mass term as the usual. Using the proper-time method \cite{Schwinger}, we show the background vectors or tensors that contribute for the effective 
electrodynamics generated by the functional integration of the fermionic fields. The letter contains just the section \ref{sec2} with the calculus of the effective lagrangian. We end the work with the conclusions, 
and one appendix showing the discrete symmetries of the model.

\vspace{0.5cm}

We adopt the Minkowski metric $\eta^{\mu\nu}=\mbox{diag}\left(+1,-1,-1,-1\right)$ with the natural units of $\hbar=c=1$ and $\epsilon_0=\mu_0=1$ (Heaviside-Lorentz units) 
in throughout this paper. The electric and magnetic fields have squared-energy dimension in this unit system. The conversion of $\mbox{MeV}^2$ to Tesla (T) in the natural system is as follows : $1 \, \mbox{T} =  5.92 \times 10^{-11} \, \mbox{MeV}^2$, respectively.

%%%%%%%%%%%%%%%%%%%%%%%%%%%%%%%%%%%%%%%%%%%%%%%%%%%%%%%%%%%%%%%%%%%%%%%%%%%%%%%%%%%%%%%%%%%%%%%%%%%%%%%%%%%%%%%%%%%%%%%%%%%%%%%%%%%%%%%%%%%%%%%%%%%%%%%%%%%%%%%%%%%%%%%%%%%%%%%%%%%%%%%%%%%%%%%%%%%%%%%%%%%%%%%%%%%%%%%%%%%%%%%%%%%%%%%%%%%%%%%%%%%%%%%%%%%%%%%%%%%%%%%%%%%%%%%%%%%%%%%%%%%%%%%%%%%%%%%%%%%%%%%%%%%%%%%%%%%%%%%%%%%%%%%%%%%%%%%%%%%%%%%%%%%%%%%%%%%%%%%%%%%%%%%%%%%%%%%%%%%%%%%%%%%%%%%%%%%%%%%%%%%%%%%%%%%%%%%%%%%%%%%%%%%%%%%%%%%%%%%%%%%%%%%%%%%%%%%%%%%%%%%%%%%%%%%%%%%%%%%%%%%%%%%%%%%%%%%%%%%%%%%%%%%%%%%%%%%%%%%%%%%%%%%%%%%%%%%%%%%%%%%%%%%%%%%%%%%%%%%%%%%%%%%%%%%%%%%%%%%%%%%%%%%%%%%%%%%%%%%%%%%%%%%%%%%%%%%%%%%%%%%%%%%%%%%%%%%%%%%%%%%%%%%%%%%%%%%%%%%%%%%%%%%%%%%%%%%%%%%%%%%%%%%%%%%%%%%%%%%%%%%%%%%%%%%%%%%%%%%%%%%%%%%%%%%%%%%%%%%%%%%%%%%%

%
\section{Effective action for fermions in the presence of LSV parameters}
\label{sec2}

The most general QED with LSV in the fermion sector is governed by the action \cite{Colladay2} :
\begin{eqnarray}\label{ActionLSV}
S_{QED}^{LSV} =  \int d^{4}x\left[ \, -\frac{1}{4} \, F_{\mu\nu}F^{\mu\nu} + \overline{\psi} \, i \, \Gamma^{\nu} D_{\nu}\psi-\overline{\psi} \,  M \, \psi \, \right] \; ,
\end{eqnarray}
where $\Gamma^{\nu}$ and $M$ are matrices defined by
\begin{subequations}
\begin{eqnarray}
\Gamma^{\nu} &=& \gamma^{\nu}+c^{\mu\nu}\gamma_{\mu}+d^{\mu\nu}\,\gamma_{\mu}\,\gamma_{5}+e^{\nu}\,\mathds{1}
+i\,f^{\nu}\,\gamma_{5}+\frac{1}{2} \, g^{\lambda\mu\nu}\,\sigma_{\lambda\mu} \; ,
\label{Gammanu}
\\
M &=& m\,\mathds{1}+a_{\mu}\gamma^{\mu}+b_{\mu}\gamma^{\mu}\gamma_{5}+\frac{1}{2}\,H_{\mu\nu}\,\sigma^{\mu\nu} \; ,
\label{Mop}
\end{eqnarray}
\end{subequations}
for the background vectors and tensors $\left\{ \, a^{\mu} \, , \, b^{\mu} \, , \, H^{\mu\nu} \, , \, c^{\mu\nu} \, , \, d^{\mu\nu} \, , \, e^{\nu} \, , \, f^{\nu} \, , \, g^{\lambda\mu\nu} \, \right\}$ that violate the Lorentz symmetry in the model. All these quantities are considered constants and uniforms throughout this paper. We also consider the $c^{\mu\nu}$-tensor traceless since the trace part of $c^{\mu\nu}$ can always be absorbed by definition of the $e$ coupling constant \cite{Brett}.     
In (\ref{Gammanu}), $\gamma^{\mu}=(\gamma^{0},\gamma^{i})$ are the Dirac matrices, $\gamma_{5}=i\,\gamma^{0}\gamma^{1}\gamma^{2}\gamma^{3}$, $\sigma^{\mu\nu}=i \left[\gamma^{\mu},\gamma^{\nu}\right]/2$, and in (\ref{Mop}), $m$ is the electron mass of $0.5$ MeV. In the action (\ref{ActionLSV}), $D_{\mu} = \partial_{\mu}+i \, e \, A_{\mu}$ is the covariant derivative operator, and $F_{\mu\nu} =  \partial_{\mu}A_{\nu}-\partial_{\nu}A_{\mu}$ is the strength field tensor for the EM field, as usual.
Under the discrete CPT-symmetry, the action (\ref{ActionLSV}) is invariant if $b^{\mu}$ and $f^{\nu}$ are pseudo-vectors, and also if $d^{\mu\nu}$ is a pseudo-tensor.
We show these properties in more details in the appendix A.

In this paper, we make $a^{\mu}=b^{\mu}=H^{\mu\nu}=0$, and we investigate the effective action for the case with all the background vectors and tensors are non-nulls 
from (\ref{Gammanu}). Using the known technics from QFT, we make the functional integration over the fermionic fields :
\begin{eqnarray}
\int D\psi \, D\overline{\psi} \, e^{i\int d^{4}x \, \overline{\psi} \left( \, i \, \Gamma^{\nu}D_{\nu}-m\,\right)\psi} = \det\left( \, i \, \Gamma^{\nu}D_{\nu}-m \, \right) = e^{\textrm{Tr}\ln\left( \, i \, \Gamma^{\nu}D_{\nu}-m \, \right)}=e^{i\,\Gamma^{\left(1\right)}} \; ,
\end{eqnarray}
where the effective action at one-loop is given by
\begin{eqnarray}\label{Actioneff1}
\Gamma^{\left(1\right)} =  -\,i\,\textrm{Tr}\ln\left(\,i\,\Gamma^{\nu}D_{\nu}-m\,\right) \; .
\end{eqnarray}
This is the quantum correction to the Maxwell action in (\ref{ActionLSV}), where $\mbox{Tr}$ symbols the trace operation in the coordinate space, and also over 
all the Dirac matrices that appear inside of $\Gamma^{\nu}$. We use the properties of $\gamma_{5}$-matrix, such as $\gamma_{5}^{2}=\mathds{1}$, $\left\{ \, \gamma_{5} \, ,\, \gamma_{\mu} \, \right\} =0$, and the cyclic property of the trace, thus the effective action (\ref{Actioneff1}) is 
%
%\begin{eqnarray}
%S_{Maxwell} & = & \int d^{4}x\left(-\frac{1}{4}F_{\mu\nu}F^{\mu\nu}\right).
%\end{eqnarray}
%
\begin{eqnarray}
\Gamma^{\left(1\right)} &=& -\,i\,\textrm{Tr}\ln\left[\,\left( \, i\,\Gamma^{\nu}D_{\nu}-m \, \right) \gamma_{5}^2 \,\right]  
\nonumber \\
&&
\hspace{-0.4cm}
= -\, i \, \textrm{Tr}\ln\left[ i\left(-\gamma^{\nu}-c^{\mu\nu}\gamma_{\mu}-d^{\mu\nu}\gamma_{\mu}\gamma_{5}+e^{\nu}\,\mathds{1}+i\,f^{\nu}\gamma_{5}
+\frac{1}{2} \, g^{\lambda\mu\nu}\sigma_{\lambda\mu}\right)D_{\nu}-m\right] \; .
\end{eqnarray}
Thereby, we can write the effective action as
\begin{eqnarray}
\Gamma^{\left(1\right)} & = & -\frac{i}{2} \, \textrm{Tr}\ln\left[ \, i\left(\gamma^{\nu}+c^{\mu\nu}\gamma_{\mu}+d^{\mu\nu}\gamma_{\mu}\gamma_{5}+e^{\nu}\,\mathds{1}+i\,f^{\nu}\,\gamma_{5}+\frac{1}{2} \, g^{\lambda\mu\nu}\sigma_{\lambda\mu}\right)D_{\nu}-m\,\right]
\nonumber \\
&& 
-\frac{i}{2} \,\textrm{Tr}\ln\left[ i\left(-\gamma^{\nu}-c^{\mu\nu}\gamma_{\mu}-d^{\mu\nu}\gamma_{\mu}\gamma_{5}+e^{\nu}\,\mathds{1}+i\,f^{\nu}\,\gamma_{5}+\frac{1}{2} \, g^{\lambda\mu\nu}\sigma_{\lambda\mu}\right)D_{\nu}-m\right]
\nonumber \\
&&
=-\frac{i}{2} \, \textrm{Tr}\ln\left\{ \left[ \, i\left(\gamma^{\nu}+c^{\mu\nu}\gamma_{\mu}+d^{\mu\nu}\gamma_{\mu}\gamma_{5}+e^{\nu}\,\mathds{1}+i\,f^{\nu}\,\gamma_{5}+\frac{1}{2} \, g^{\lambda\mu\nu}\sigma_{\lambda\mu}\right)D_{\nu}-m\,\right]\right.
\nonumber \\
 && 
 \times
 \left.\left[i\left(-\gamma^{\alpha}-c^{\beta\alpha}\gamma_{\beta}-d^{\beta\alpha}\gamma_{\beta}\gamma_{5}+e^{\alpha}\,\mathds{1}+i\,f^{\alpha}\,\gamma_{5}+\frac{1}{2} \, g^{\kappa\beta\alpha}\sigma_{\kappa\beta}\right)D_{\alpha}-m\right]\right\} \; .
\end{eqnarray}
We consider just the linear terms in the background vectors/tensors, so inside the logarithm function, we have
\begin{eqnarray}
\Gamma^{\left(1\right)} \simeq - \, \frac{i}{2} \, \textrm{Tr}\ln(A+B) \; , 
\end{eqnarray}
where $A=D^{2}+\frac{e}{2}\,\sigma^{\mu\nu}F_{\mu\nu}+m^{2}$, and $B$ is set by the matrix
\begin{eqnarray}\label{B}
B &=& 
c^{\mu\nu} \, \slashed{D} \, \gamma_{\mu} \, D_{\nu}+c^{\mu\nu} \, \gamma_{\mu} \, D_{\nu} \, \slashed{D}
+ d^{\mu\nu} \slashed{D} \, \gamma_{\mu} \gamma_{5} \, D_{\nu}+d^{\mu\nu} \gamma_{\mu} \, \gamma_{5} \, D_{\nu} \, \slashed{D}
\nonumber \\
&&
-\,i \, e \, e^{\nu} \gamma^{\mu} F_{\mu\nu}+i\,f^{\nu}\gamma_{5}\,\gamma^{\mu}\left\{ \, D_{\mu} \, , \, D_{\nu} \, \right\}
+\frac{1}{2} \, g^{\lambda\mu\nu}\sigma_{\lambda\mu}D_{\nu} \, \slashed{D}
-\frac{1}{2} \, g^{\lambda\mu\nu} \, \slashed{D} \, \sigma_{\lambda\mu} D_{\nu}
\nonumber \\
&&
-i\,2m\left(e^{\nu}\,\mathds{1}+i\,f^{\nu}\,\gamma_{5}+\frac{1}{2} \, g^{\lambda\mu\nu} \sigma_{\lambda\mu}\right)D_{\nu} \; .
\end{eqnarray} 
For very small coefficients in $B$, the effective action is approximated by 
\begin{eqnarray}\label{Gamma1QEDLSV}
\Gamma^{\left(1\right)} \simeq - \, \frac{i}{2} \, \textrm{Tr}\ln(A+B) \simeq - \frac{i}{2} \, \textrm{Tr}\ln\left(A\right)-\frac{i}{2} \, \textrm{Tr}\left(B\,A^{-1}\right) \; .
\end{eqnarray}
Notice that $\Gamma^{\left(1\right)} \simeq \Gamma_{QED}^{\left(1\right)}+\Gamma_{LSV}^{\left(1\right)}$, where $\Gamma_{QED}^{\left(1\right)}$ is the
effective action for the QED, and $\Gamma_{LSV}^{\left(1\right)}$ is the effective action for the LSV from the second term of (\ref{Gamma1QEDLSV}). The 
renormalized effective lagrangian for the QED has the result \cite{EulerHeisenberg}
\begin{eqnarray}\label{LQED}
{\cal L}_{QED}^{(1)}= \frac{1}{32\pi^2}\int_{0}^{\infty} \frac{ds}{s} \, e^{ism^2} \left[ \, 4e^2\,{\cal G}\,\coth(esa)\,\cot(esb)-\frac{4}{s^2}+\frac{8e^2}{3}\,{\cal F} \, \right] \; ,\label{Euler_Heisenberg_action}
\end{eqnarray}
where $a$ and $b$ are functions of the Lorentz-invariants, namely, $a = \sqrt{\sqrt{\mathcal{F}^{2}+\mathcal{G}^{2}}-\mathcal{F}}$ and
$b = \sqrt{\sqrt{\mathcal{F}^{2}+\mathcal{G}^{2}}+\mathcal{F}}$, with $\mathcal{F} = \frac{1}{4} \, F_{\mu\nu}F^{\mu\nu}$ and
$\mathcal{G} = \frac{1}{4} \, F_{\mu\nu}\tilde{F}^{\mu\nu}$. Notice that \eqref{Euler_Heisenberg_action} is the Euler-Heisenberg effective lagrangian. 
For weak EM fields, the expression (\ref{LQED}) reduces to
\begin{eqnarray}
{\cal L}_{QED}^{(1)}\simeq \frac{e^4}{360\pi^2m^4} \left(4\,{\cal F}^2+7\,{\cal G}^2 \right) \; .
\end{eqnarray}
Returning to the LSV effective action, we use the Schwinger integral for $A^{-1}$ in (\ref{Gamma1QEDLSV})
\begin{eqnarray}
A^{-1} = - \, i \, \int_{0}^{\infty}ds \, e^{i s A} \; ,
\end{eqnarray}
in which $s$ is the proper-time parameter, the LSV effective action is read as 
\begin{eqnarray}\label{Gamma1LSV}
\Gamma_{LSV}^{\left(1\right)}\simeq - \, \frac{1}{2} \,\textrm{Tr} \int_{0}^{\infty}ds \, B \, e^{isA} \; .
\end{eqnarray}
Inside the trace (\ref{Gamma1LSV}), all the terms of the second line in (\ref{B}) are null due to the trace of odd number in the $\gamma^{\mu}$-Dirac matrices.
Thus, we write (\ref{Gamma1LSV}) as $\Gamma_{LSV}^{\left(1\right)}=\Gamma_{LSV-1}^{\left(1\right)}+\Gamma_{LSV-2}^{\left(1\right)}+\Gamma_{LSV-3}^{\left(1\right)}$, 
where
\begin{subequations}
\begin{eqnarray}\label{Gamma1LSV1}
\Gamma_{LSV-1}^{\left(1\right)}  &=& -\frac{1}{2}\int_{0}^{\infty}ds \, e^{ism^{2}} \, \textrm{Tr}\left[\left(c^{\mu\nu}\left\{ \, D_{\mu} \, , \, D_{\nu}\right\} 
-e\,c^{\mu\nu} \sigma_{\mu\rho} \, F^{\rho}_{\;\;\;\nu}\right)e^{isD^{2}}e^{\frac{ise}{2}\sigma_{\alpha\beta}F^{\alpha\beta}}\right] \; ,
\label{Gamma1LSV1}
\\
\Gamma_{LSV-2}^{\left(1\right)}  &=&  -\frac{i}{2}\int_{0}^{\infty}ds \, e^{ism^{2}} \, \textrm{Tr}\left[\left(e\,\gamma_{5}\,d^{\mu\nu}F_{\mu\nu}+\gamma_{5}\,d^{\mu\nu}\,\sigma_{\mu\rho} \left\{ \, D^{\rho} \, , \, D_{\nu} \, \right\} \right)e^{isD^{2}}e^{\frac{ise}{2}\,\sigma_{\alpha\beta}F^{\alpha\beta}}\right] \; ,
\label{Gamma1LSV2}
\\
\Gamma_{LSV-3}^{\left(1\right)}  &=&  i\,m \int_{0}^{\infty}ds \, e^{ism^{2}} \, \textrm{Tr}\left[\, \left(e^{\nu}\,\mathds{1}+i\,f^{\nu}\gamma_{5}+\frac{1}{2} \, g^{\nu\lambda\mu}\sigma_{\lambda\mu}\right)e^{\frac{ise}{2}\,\sigma_{\alpha\beta}F^{\alpha\beta}} \left(D_{\nu}\,e^{isD^{2}}\right) \, \right] \; .
\label{Gamma1LSV3}
\end{eqnarray}
\end{subequations}
In all these integrals, we are considering the case in which the EM tensor $F^{\mu\nu}$ is constant and uniform, thus, we can calculate the integrals separately.    
Explicitly, the traces of (\ref{Gamma1LSV1})-(\ref{Gamma1LSV3}) are given by
\begin{subequations}
\begin{eqnarray}
&&
\textrm{Tr}\left[\left(c^{\mu\nu}\left\{ \, D_{\mu} \, , \, D_{\nu}\right\} 
-e\,c^{\mu\nu} \sigma_{\mu\rho} \, F^{\rho}_{\;\;\,\nu}\right)e^{isD^{2}}e^{\frac{ise}{2}\sigma_{\alpha\beta}F^{\alpha\beta}}\right]
\nonumber \\
&&
=\int d^{4}x \, \left\{ \,
c^{\mu\nu} \, \langle x | \left\{ \, D_{\mu} \, , \, D_{\nu}\right\} \, e^{isD^{2}}  | x \rangle \, \mbox{tr}\left[ \, e^{\frac{ise}{2}\sigma_{\alpha\beta}F^{\alpha\beta}}\right] 
- \, e\,c^{\mu\nu}F^{\rho}_{\;\;\,\nu} \, \mbox{tr}\left[ \sigma_{\mu\rho} \, e^{\frac{ise}{2}\sigma_{\alpha\beta}F^{\alpha\beta}} \right] 
\langle x | e^{isD^{2}}  | x \rangle \, \right\} \; ,
\label{Tr1}
\\
&&
\textrm{Tr}\left[\left(e\,\gamma_{5}\,d^{\mu\nu}F_{\mu\nu}+\gamma_{5}\,d^{\mu\nu}\,\sigma_{\;\;\;\nu}^{\rho}\left\{ \, D_{\mu} \, , \, D_{\rho} \, \right\} \right)e^{isD^{2}}e^{\frac{ise}{2}\sigma_{\alpha\beta}F^{\alpha\beta}}\right]
\nonumber \\
&&
=\int d^4x \, \left\{ \,  e \,d^{\mu\nu}F_{\mu\nu} \, 
\mbox{tr}\left[\gamma_{5}\,e^{\frac{ise}{2}\,\sigma_{\alpha\beta}F^{\alpha\beta}}\right] \langle x | e^{isD^{2}}  | x \rangle
+ d^{\mu\nu} \, \mbox{tr}\left[ \gamma_{5} \, \sigma_{\mu\rho} \, e^{\frac{ise}{2}\,\sigma_{\alpha\beta}F^{\alpha\beta}} \right]
\langle x | \left\{ \, D^{\rho} , D_{\nu} \, \right\} \, e^{isD^{2}}  | x \rangle \, \right\} \; ,
\label{Tr2}
\\
&&
\textrm{Tr}\left[\, \left(e^{\nu}\,\mathds{1}+i\,f^{\nu}\gamma_{5}+\frac{1}{2} \, g^{\nu\lambda\mu}\sigma_{\lambda\mu}\right) e^{\frac{ise}{2}\,\sigma_{\alpha\beta}F^{\alpha\beta}} \left( D_{\nu} \, e^{isD^{2}} \right) \, \right]=
\nonumber \\
&&
=\int d^4x \, \mbox{tr}\left[ \left(e^{\nu}\,\mathds{1}+i\,f^{\nu}\gamma_{5}+\frac{1}{2} \, g^{\nu\lambda\mu}\sigma_{\lambda\mu}\right) e^{\frac{ise}{2}\,\sigma_{\alpha\beta}F^{\alpha\beta}} \right]
\langle x | \, D_{\nu} \, e^{isD^{2}} \, | x \rangle \; ,
\label{Tr3}
\end{eqnarray}
\end{subequations}
respectively, where $\mbox{tr}$ now denotes the trace operation over the Dirac matrices. The last trace (\ref{Tr3}) is null due to the result for the kernel with just one $D_{\nu}$-operator
\begin{eqnarray}
\langle x | \, D_{\nu} \, e^{isD^{2}} \, | x \rangle = 0 \; ,
\end{eqnarray}
that is valid when $F^{\mu\nu}$ is constant and uniform. Consequently, the action (\ref{Gamma1LSV3}) is null:
\begin{eqnarray}
\Gamma_{LSV-3}^{\left(1\right)}=0 \; .
\end{eqnarray}
The traces (\ref{Tr1}) and (\ref{Tr2}) depend on the kernel \cite{McArthur}:
\begin{eqnarray}\label{kernel}
\langle x|e^{isD^{2}} |x\rangle = \frac{-\,i}{16\pi^2 s^2} \, \mbox{det}^{1/2}\left[\frac{es\mathbb{F}}{\sin\left(es\mathbb{F}\right)}\right] \; ,
\end{eqnarray}
in which $\mathbb{F}$ symbols the matrix of the $F^{\mu\nu}$-tensor. Since the determinant of a $n \times n$-matrix is the product of the eigenvalues, the result 
(\ref{kernel}) is given by   
\begin{eqnarray}\label{kernelresult}
\langle x|e^{isD^{2}}|x\rangle = \frac{-\,i\,e^2}{16\pi^2} \, \frac{a \, b}{\sinh\left(esa\right)\sin(esb)} \; .
\end{eqnarray}
The extension of the kernel (\ref{kernel}) that is useful for us is \cite{Schwinger}
\begin{equation}
\left\langle x\right|\left\{ D_{\mu},D_{\nu}\right\} e^{isD^{2}}\left|x\right\rangle = \frac{1}{16\pi^2 s^2}\left[e\,\mathbb{F}\cot\left(es\,\mathbb{F}\right)\right]_{\mu\nu} \mbox{det}^{1/2}\left[\frac{es\,\mathbb{F}}{\sin\left(es\mathbb{F}\right)}\right]
=\frac{e^3}{16\pi^2} \, \frac{a \, b \, \left[\mathbb{F}\cot\left(es\mathbb{F}\right)\right]_{\mu\nu}}{\sinh\left(esa\right)\sin(esb)}   \; .
\end{equation}  
The traces over the Dirac matrix that we need here are given by 
\begin{subequations}
\begin{eqnarray}
\mbox{tr}\left[e^{\frac{ise}{2}\,\sigma_{\mu\nu}F^{\mu\nu}}\right] &=&  2 \left[ \, \cos\left( e s \, \sqrt{{\cal F}+i\,{\cal G}} \right) +\cos\left( e s \, \sqrt{{\cal F}-i\,{\cal G}} \right) \, \right] \; ,
\\
\mbox{tr}\left[ \sigma_{\mu\rho} \, e^{\frac{ise}{2}\sigma_{\alpha\beta}F^{\alpha\beta}} \right] &=& \, 2i\,F_{\mu\rho} \left[ \frac{\sin\left(es\,\sqrt{ {\cal F}+i\,{\cal G} } \right)}{\sqrt{{\cal F}+i\,{\cal G}}} +\frac{\sin\left(es\,\sqrt{ {\cal F}-i\,{\cal G} } \right)}{\sqrt{{\cal F}-i\,{\cal G}}} \right]
\nonumber \\
&&
+2 \, \tilde{F}_{\mu\rho} \left[ \frac{\sin\left(es\,\sqrt{ {\cal F}+i\,{\cal G} } \right)}{\sqrt{{\cal F}+i\,{\cal G}}} -\frac{\sin\left(es\,\sqrt{ {\cal F}-i\,{\cal G} } \right)}{\sqrt{{\cal F}-i\,{\cal G}}} \right] \; ,
\\
\mbox{tr}\left[\gamma_{5} \, e^{\frac{ise}{2}\sigma_{\mu\nu}F^{\mu\nu}}\right] &=& 2 \left[ \, \cos\left( e s \, \sqrt{{\cal F}+i\,{\cal G}} \right) -\cos\left( e s \, \sqrt{{\cal F}-i\,{\cal G}} \right) \, \right] \; ,
\\
\mbox{tr} \left[\gamma_{5} \, \sigma_{\mu\rho} \, e^{\frac{ise}{2}\sigma_{\alpha\beta}F^{\alpha\beta}}\right] &=&  2 i \, F_{\mu\rho} \left[ \frac{\sin\left(es\,\sqrt{ {\cal F}+i\,{\cal G} } \right)}{\sqrt{{\cal F}+i\,{\cal G}}} -\frac{\sin\left(es\,\sqrt{ {\cal F}-i\,{\cal G} } \right)}{\sqrt{{\cal F}-i\,{\cal G}}} \right]
\nonumber \\
&&
+2\,\tilde{F}_{\mu\rho} \left[ \frac{\sin\left(es\,\sqrt{ {\cal F}+i\,{\cal G} } \right)}{\sqrt{{\cal F}+i\,{\cal G}}} +\frac{\sin\left(es\,\sqrt{ {\cal F}-i\,{\cal G} } \right)}{\sqrt{{\cal F}-i\,{\cal G}}} \right]
\; .
\end{eqnarray}
\end{subequations}
Using all these results, the effective action (\ref{Gamma1LSV1}) is written as $\Gamma_{LSV-1}^{(1)} = \int d^4x \, {\cal L}^{(1)}_{LSV-1}$, where the effective lagrangian ${\cal L}^{(1)}_{LSV-1}$ is
\begin{eqnarray}
{\cal L}^{(1)}_{LSV-1} &=& -\frac{e}{16\pi^2} \int_{0}^{\infty} \frac{ds}{s^2} \, e^{ism^2} \left\{ \, c^{\mu\nu} \left[ \mathbb{F}\cot\left(es\mathbb{F}\right) \right]_{\mu\nu} 
\left[ \, \cos \left( e s \, \sqrt{{\cal F}+i\, {\cal G}} \right)+\cos \left( e s \, \sqrt{{\cal F}-i\, {\cal G}} \right) \, \right] 
\right.
\nonumber \\
&&
\left.
- c^{\mu\nu} F_{\mu\rho} F^{\rho}_{\;\;\;\nu} \left[ \, \frac{\sin\left(es\,\sqrt{ {\cal F}+i\,{\cal G} } \right)}{\sqrt{{\cal F}+i\,{\cal G}}} +\frac{\sin\left(es\,\sqrt{ {\cal F}-i\,{\cal G} } \right)}{\sqrt{{\cal F}-i\,{\cal G}}} \, \right] 
\right.
\nonumber \\
&&
\left.
+ i\, c^{\mu\nu} \tilde{F}_{\mu\rho} F^{\rho}_{\;\;\;\nu} \left[ \, \frac{\sin\left(es\,\sqrt{ {\cal F}+i\,{\cal G} } \right)}{\sqrt{{\cal F}+i\,{\cal G}}} -\frac{\sin\left(es\,\sqrt{ {\cal F}-i\,{\cal G} } \right)}{\sqrt{{\cal F}-i\,{\cal G}}}  \right] 
\, \right\} \mbox{det}^{1/2}\left[\frac{es\mathbb{F}}{\sin\left(es\mathbb{F}\right)} \, \right] \; , \label{lsv_1}
\end{eqnarray}
and similarly, the effective lagrangian ${\cal L}^{(1)}_{LSV-2}$ is
\begin{eqnarray}
{\cal L}^{(1)}_{LSV-2} &=& -\frac{e}{16\pi^2} \int_{0}^{\infty} \frac{ds}{s^2} \, e^{ism^2} \left\{ \, d^{\mu\nu}F_{\mu\nu} \left[ \, \cos \left( e s \, \sqrt{{\cal F}+i\, {\cal G}} \right)-\cos \left( e s \, \sqrt{{\cal F}-i\, {\cal G}} \right) \, \right] 
\right.
\nonumber \\
&&
\left.
- d^{\mu\nu}F_{\mu\rho} \left[ \mathbb{F}\cot\left(es\mathbb{F}\right) \right]^{\rho}_{\;\;\;\nu}
\left[ \, \frac{\sin\left(es\,\sqrt{ {\cal F}+i\,{\cal G} } \right)}{\sqrt{{\cal F}+i\,{\cal G}}} -\frac{\sin\left(es\,\sqrt{ {\cal F}-i\,{\cal G} } \right)}{\sqrt{{\cal F}-i\,{\cal G}}} \right] 
\right.
\nonumber \\
&&
\left.
+i\,d^{\mu\nu}\tilde{F}_{\mu\rho} \left[ \mathbb{F}\cot\left(es\mathbb{F}\right) \right]^{\rho}_{\;\;\;\nu} \left[ \, \frac{\sin\left(es\,\sqrt{ {\cal F}+i\,{\cal G} } \right)}{\sqrt{{\cal F}+i\,{\cal G}}} +\frac{\sin\left(es\,\sqrt{ {\cal F}-i\,{\cal G} } \right)}{\sqrt{{\cal F}-i\,{\cal G}}} \right] \, \right\} \mbox{det}^{1/2}\left[\frac{es\mathbb{F}}{\sin\left(es\mathbb{F}\right)}\right]  \; .\label{lsv_2}
\end{eqnarray}
For a weak EM field, we introduce a $s_0$-parameter ($s_0^{-1/2}$ has mass dimension) to regularize the $s$-integral in the lower limit. 
The results for the effective lagrangian are given by  
\begin{eqnarray}\label{LSV1eff}
{\cal L}^{(1)}_{LSV-1}(s_0) &\simeq & 
%-\frac{c}{8\pi^2}\left( \frac{1}{2s_0^2}+\frac{im^2}{s_0}  \right) 
%+ \frac{e^2}{48\pi^2} \left[ \, \gamma_{E}+ \ln(-im^2s_0) \,\right]
%\left( c\,{\cal F} +8\,c^{\mu\nu}F_{\mu\alpha}F^{\alpha}_{\;\;\,\nu} \right)
\frac{e^2}{6\pi^2} \left[ \, \gamma_{E}+ \ln(-im^2s_0) \,\right]
c^{\mu\nu}F_{\mu\alpha}F^{\alpha}_{\;\;\,\nu}
%\nonumber \\
%&&
%\hspace{-0.5cm}
%-\frac{c\, e^4}{2880\pi^2m^4} \left( 17 \,{\cal F}^2 +11 \, {\cal G}^2 \right) 
-\frac{e^4}{72\pi^2m^4} \, c^{\mu\nu}F_{\mu\alpha}F^{\alpha}_{\;\;\,\nu}\,{\cal F}
+\frac{e^4}{48\pi^2m^4} \, c^{\mu\nu}\tilde{F}_{\mu\alpha}F^{\alpha}_{\;\;\,\nu}\,{\cal G} \; , \label{lsv_weak_field_1}
\end{eqnarray}
and
\begin{eqnarray}\label{LSV2eff}
{\cal L}^{(1)}_{LSV-2}(s_0) &\simeq& 
-\frac{i\,e}{8\pi^2} \, d^{\mu\nu}\tilde{F}_{\mu\nu} \left[ \, \frac{1}{s_0}+im^2(1-\gamma_{E})-im^2\ln(-im^2s_0) \, \right]
-\frac{e^3}{24\pi^2m^2}\,d^{\mu\nu}F_{\mu\nu} \, {\cal G}
\nonumber \\
&&
\hspace{-0.5cm}
+\frac{e^3}{48\pi^2m^2} \, d^{\mu\nu}\tilde{F}_{\mu\nu}\,{\cal F}
-\frac{e^3}{24\pi^2m^2}\,d^{\mu\nu}\tilde{F}_{\mu\alpha}F^{\alpha\rho}F_{\rho\nu} 
\; ,\label{lsv_weak_field_2}
\end{eqnarray}
where $\gamma_{E}=0.577$ is the Euler-Mascheroni constant. The terms in the first line from (\ref{LSV1eff}) and (\ref{LSV2eff}) are divergent in the limit $s_{0} \rightarrow 0$. 
In (\ref{LSV1eff}), we need to add counterterms to cancel these divergences (see appendix A). In (\ref{LSV2eff}), the divergent term is interpreted as a surface term inside the correspondent action, {\it i.e.}, this term is null and can be discarded in the effective lagrangian for weak fields. Therefore, meeting the 
physical terms with the first quantum corrections leads to the result
\begin{eqnarray}\label{LeffNL}
{\cal L}_{eff} &=& -\frac{1}{4} \, F_{\mu\nu}F^{\mu\nu}+\frac{e^4}{360\pi^2m^4} \left(4\,{\cal F}^2+7\,{\cal G}^2 \right)
%-\frac{c\, e^4}{2880\pi^2m^4} \left( 17 \,{\cal F}^2 +11 \, {\cal G}^2 \right) 
%\nonumber \\
%&&
-\frac{e^4}{72\pi^2m^4} \, c^{\mu\nu}F_{\mu\alpha}F^{\alpha}_{\;\;\,\nu}\,{\cal F}
+\frac{e^4}{48\pi^2m^4} \, c^{\mu\nu}\tilde{F}_{\mu\alpha}F^{\alpha}_{\;\;\,\nu}\,{\cal G}
\nonumber \\
&&
-\frac{e^3}{24\pi^2m^2}\,d^{\mu\nu}F_{\mu\nu} \, {\cal G}
+\frac{e^3}{48\pi^2m^2} \, d^{\mu\nu}\tilde{F}_{\mu\nu}\,{\cal F}
-\frac{e^3}{24\pi^2m^2}\,d^{\mu\nu}\tilde{F}_{\mu\alpha}F^{\alpha\rho}F_{\rho\nu} \; .
\end{eqnarray}
We obtain a non-linear electrodynamics with quartic terms in the EM field associated with the $c^{\mu\nu}$-terms, and with third terms in 
$F^{\mu\nu}$ for the sector of $d^{\mu\nu}$-tensor. The terms in $c^{\mu\nu}$ in the second line are similar to that found in the ref. \cite{Araujo}. 
As we expect, the limits $c^{\mu\nu} \rightarrow 0$ and $d^{\mu\nu} \rightarrow 0$ recover the Euler-Heisenberg term for weak EM field. Furthermore, 
the lagrangian density (\ref{LeffNL}) preserves the CPT-symmetry, since that $d^{\mu\nu}$ is a pseudo-tensor, even under charge conjugation (see appendix A).   
Considering the range of values for the $c^{\mu\nu}$-components in $|c| \sim 10^{-18}-10^{-13}$, see the ref. \cite{tables} (table D6),  
%
%For a Magnetar, the magnetic field is of order $|{\bf B}| \simeq 10^{11} \, \mbox{T}$, and using 
%
the quartic terms in $F^{\mu\nu}$ of the effective lagrangian are relevant if the magnetic field is $|{\bf B}| \sim 10^{18}-10^{21} \, \mbox{T}$.
In the case of cubic terms in $F^{\mu\nu}$, in the range $|d| \sim 10^{-17}-10^{-14}$, the estimative for the magnetic field is 
$|{\bf B}| \sim 10^{27}-10^{30} \, \mbox{T}$.

%**************************

%\section{Exploring the non-linear electrodynamics}
%\label{sec3}
%

%\section{Dispersion relations in a magnetic background field}
%\label{sec4}
%

%%%%%%%%%%%%%%%%%%%%%%%%%%%%%%%%%%%%%%%%%%%%%%%%%%%%%%%%%%%%%%%%%%%%%%%%%%%%%%%%%%%%%%%%%%%%%%%%%%%%%%%%%%%%%%%%%%%%%%%%%%%%%%%%%%%%%%%%%%%%%%%%%%%%%%%%%%%%%%%%%%%%%%%%%%%%%%%%%%%%%%%%%%%%%%%%%%%%%%%%%%%%%%%%%%%%%%%%%%%%%%%%%%%%%%%%%%%%%%%%%%%%%%%%%%%%%%%%%%%%%%%%%%%%%%%%%%%%%%%%%%%%%%%%%%%%%%%%%%%%%%%%%%%%%%%%%%%%%%%%%%%%%%%%%%%%%%%%%%%%%%%%%%%%%%%%%%%%%%%%%%%%%%%%%%%%%%%%%%%%%%%%%%%%%%%%%%%%%%%%%%%%%%%%%%%%%%%%%%%%%%%%%%%%%%%%%%%%%%%%%%%%%%%%%%%%%%%%%%%%%%%%%%%%%%%%%%%%%%%%%%%%%%%%%%%%%%%%%%%%%%%%%%%%%%%%%%%%%%%%%%%%%%%%%%%%%%%%%%%%%%%%%%%%%%%%%%%%%%%%%%%%%%%%%%%%%%%%%%%%%%%%%%%%%%%%%%%%%%%%%%%%%%%%%%%%%%%%%%%%%%%%%%%%%%%%%%%%%%%%%%%%%%%%%%%%%%%%%%%%%%%%%%%%%%%%%%%%%%%%%%%%%%%%%%%%%%%%%%%%%%%%%%%%%%%%%%%%%%%%%%%%%%%%%%%%%%%%%%%%%%%%%%%%%

\section{Conclusions}
\label{sec5}

In this letter, we have calculated the effective action at one-loop for the quantum electrodynamics in the scenario of Lorentz symmetry violation in which 
background vectors and tensors are introduced as a combination of the Dirac matrix $\left\{ \, \gamma^{\mu} \, , \, \gamma_{5} \, \right\}$ in the fermion sector, 
as $\Gamma^{\nu} = \gamma^{\nu}+c^{\mu\nu}\gamma_{\mu}+d^{\mu\nu}\,\gamma_{\mu}\,\gamma_{5}+e^{\nu}\,\mathds{1}
+i\,f^{\nu}\,\gamma_{5}+\frac{1}{2} \, g^{\lambda\mu\nu}\,\sigma_{\lambda\mu}$. We considered the case in which the background vectors/tensors 
$\left\{ \, c^{\mu\nu} \, , \, d^{\mu\nu}\, , \, e^{\nu} \, , \, f^{\nu} \, , \, g^{\lambda\mu\nu} \, \right\}$ are constants in the 
linear approximation. In the case of uniform and constant electromagnetic field, the quantities $\left\{ \, e^{\nu} \, , \, f^{\nu} \, , \, g^{\lambda\mu\nu} \, \right\}$ 
do not contribute for the effective lagrangian at one-loop and in the linear approximation, in which the effective action associated with them is null. Thereby, only the background tensors $\left\{ \, c^{\mu\nu} \, , \, d^{\mu\nu} \, \right\}$ provides information of a new physics in this stage. The result for the effective lagrangian is so obtained through the Schwinger's proper time method. In approximation of a weak electromagnetic field, the correspondent effective lagrangian generates a non-linear electrodynamics that is quartic in the EM field strength tensor $(F^{\mu\nu})$ for contractions with the components of $c^{\mu\nu}$-tensor, and cubic in 
$F^{\mu\nu}$ for the contractions with the $d^{\mu\nu}$-tensor. The non-linear ED preserves the CPT-symmetry since that $d^{\mu\nu}$ is a pseudo-tensor. Quartic effects similar to Euler-Heisenberg ED are obtained and also cubic interactions at tree level show the emergence of a new physics. Using the known bounds for the components of $c^{\mu\nu}$ and $d^{\mu\nu}$, we obtain roughly the estimative for magnetic fields above the scale of neutron stars and Magnetars, that is one main motivations for the study of non-linear electrodynamics. The investigation of classical ED associated with the effective lagrangian (\ref{LeffNL}) is one important perspective for the project. The study of the linearization of (\ref{LeffNL}) in the presence of an external EM field can reveal characteristics of the vacuum of this model as the birefringence effect connected to components of the background tensors $\left\{ \, c^{\mu\nu} \, , \, d^{\mu\nu} \, \right\}$. Others optical effects also can be explored in a forthcoming project.

\section*{APPENDIX A : REMARKS ON SYMMETRIES AND RENORMALIZATION}

In this work, we are considering the following physical action
\begin{eqnarray}
S^{LSV}_{QED}|_{a_\mu=b_\mu=H_{\mu\nu}=0} = \int d^{4}x\left[\, -\frac{1}{4}\,F_{\mu\nu}F^{\mu\nu}+\overline{\psi} \, i \, \Gamma^{\nu}D_{\nu}\psi-m\,\overline{\psi}\,\psi \, \right] \; , \label{S_lsv}
\end{eqnarray}
where $\Gamma^\nu$ is defined by the Eq. \eqref{Gammanu}. Notice that, this action is invariant under the $U(1)$-local gauge transformations :
\begin{eqnarray}
\psi^{\prime}\left(x\right) =  e^{ie\omega\left(x\right)}\psi\left(x\right) 
\;\; , \;\;
\overline{\psi}^{\,\prime}\left(x\right)=e^{-ie\omega\left(x\right)}\psi\left(x\right)
\;\; , \;\; 
A^{\prime}_{\mu}\left(x\right)=A_{\mu}\left(x\right)+\partial_{\mu}\omega \; ,
\end{eqnarray}
for an arbitrary $\omega$-function. In addition to the gauge symmetry, $S^{LSV}_{QED}$ is invariant with respect
to the discrete transformations $C$ (charge conjugation), $P$ (parity) and $T$ (time-reversal operation), if we consider
the proper transformations to the LSV parameters. 
\begin{enumerate}[label=\arabic*)]
\item $C$-symmetry (charge conjugation)
The charge conjugation transformations for the fields are 
\begin{eqnarray}
\psi^{c} = C \, \overline{\psi}^{\,t} \; \; , \; \;
\overline{\psi}^{\,c}=\psi^{t} \, C \; \; , \; \; 
A_{\mu}^{c}=-\,A_{\mu} \; ,
\end{eqnarray}
where, $C$ is the charge conjugation matrix, which has the following
properties : $C \, \gamma^{\mu\,t} \, C=\gamma^{\mu}$ and $C^{-1}=C^{t}=-\,C$.
In the Weyl representation of the Dirac matrices,
it follows that $C=i\,\gamma^{0}\,\gamma^{2}$. Therefore, we have transformations
\begin{gather}
c^{\mu\nu\,c}=c^{\mu\nu} 
\; , \;
d^{\mu\nu\,c}=-d^{\mu\nu}
\; , \;
e^{\mu\,c}=-e^{\mu} 
\; , \;
f^{\mu\,c}=-f^{\mu}
\; , \;
g^{\mu\nu\lambda\,c}=-g^{\mu\nu\lambda} \; . \label{charge_conjugation_transformation}
\end{gather}

%\begin{enumerate}[label=\arabic*)]
\item $P$-symmetry (parity)
Under the parity transformation ($x^{\prime\mu}=P_{\;\;\,\nu}^{\mu}\,x^{\nu}$), where $P_{\;\;\,\nu}^{\mu}=\mbox{diag}(+1,-1,-1,-1)$ is the parity tensor, 
we have
\begin{eqnarray}
\psi^{P} = \gamma^{0}\,\psi
\; \; , \; \;  
\overline{\psi}^{P}=\overline{\psi} \, \gamma^{0}
\; \; , \; \;
A^{P\,\mu}=P_{\;\;\,\nu}^{\mu} \, A^{\nu} \; ,
\end{eqnarray}
thus, we require that
\begin{gather}	
c^{\mu\nu\,P}=P_{\;\;\,\alpha}^{\mu} \, P_{\;\beta}^{\nu} \, c^{\alpha\beta} \; , \;
d^{\mu\nu\,P}=-P_{\;\;\,\alpha}^{\mu}\,P_{\;\;\,\beta}^{\nu}\,d^{\alpha\beta}
\; , \;
e^{\mu\,P}=P_{\;\;\,\alpha}^{\mu} \, e^{\alpha} 
\; , \;
\nonumber \\	
f^{\mu\,P}=-\,P_{\;\;\,\alpha}^{\mu} \, f^{\alpha}
\; , \;
g^{\mu\nu\lambda\,P}=P_{\;\;\,\alpha}^{\mu} \, P_{\;\;\,\beta}^{\nu} \, P_{\;\;\,\gamma}^{\lambda} \, g^{\alpha\beta\gamma} \; .
\end{gather}

\item $T$-symmetry (time-reversal operation) 
For the spinors and the gauge field, the $T$ transformations ($x^{\prime\mu}=T_{\;\;\,\nu}^{\mu}\,x^{\nu}$) with $T_{\;\;\,\nu}^{\mu}=\mbox{diag}(-1,+1,+1,+1)$
are
\begin{eqnarray}
\psi^{T} = \gamma^{1} \, \gamma^{3} \, \psi^{*}
\; \; , \; \;
\overline{\psi}^{T}=-\,\overline{\psi}^{*} \, \gamma^{1} \, \gamma^{3}
\; \; , \; \; 
A^{T\,\mu}=T_{\;\;\,\nu}^{\mu} \, A^{\nu} \; .
\end{eqnarray}
Therefore, it follows that
\begin{gather}	
c^{\mu\nu\,T}=T_{\;\;\,\alpha}^{\mu} \, T_{\;\;\,\beta}^{\nu} \, c^{\alpha\beta} 
\; , \;
d^{\mu\nu\,T}=-\,T_{\;\;\,\alpha}^{\mu} \, T_{\;\;\,\beta}^{\nu} \, d^{\alpha\beta}
\; , \;
e^{\mu\,T}=T_{\;\;\,\alpha}^{\mu} \, e^{\alpha} \; ,
\nonumber \\	
f^{\mu\,T}=-\,T_{\;\;\,\alpha}^{\mu} \, f^{\alpha} 
\; , \;
g^{\mu\nu\lambda\,T}=T_{\;\;\,\alpha}^{\mu} \, T_{\;\;\,\beta}^{\nu} \, T_{\;\;\,\gamma}^{\lambda} \, g^{\alpha\beta\gamma} \; .
\end{gather}
\end{enumerate}
By analyzing the expressions \eqref{lsv_1} and \eqref{lsv_2}, one can verify that these discrete symmetries are preserved at the one-loop level in the approximations we are considering. In particular, the divergent part, see eqs. \eqref{lsv_weak_field_1} and \eqref{lsv_weak_field_2}, preserves $C$, $P$ and $T$-symmetries and it is local. 
Therefore, we can renormalize the effective action by introducing a local invariant counterterm, as we already pointed out in Section \ref{sec2}. Fortunately, in the linear approximation for the LSV-breaking tensors, due to these discrete symmetries, the gauge-invariance, and the tensorial structure, only one counterterm is necessary. Since surface terms, such as $d_{\mu\nu}\tilde{F}^{\mu\nu}$, do not contribute, the only counterterm required is :
\begin{gather}
%a_1 \, c \, F_{\mu\nu} \, F^{\mu\nu}
%\; \; , \; \;
a \, c^{\mu}_{\;\;\,\nu} \, F_{\mu\lambda} \, F^{\lambda\nu}
\; \; , \; \;
%a_3 \, c \, m^4 \; ,
\end{gather} 
where $a$ is a constant fixed by renormalization condition. Any other gauge-invariant local counterterm of dimension four involving two strength field tensors $F_{\mu\nu}$, and a LSV tensor is forbidden by the charge conjugation \eqref{charge_conjugation_transformation}.
If we go beyond the linear approximation for the LSV parameters, to renormalize the theory defined by \eqref{S_lsv}, we must introduce an infinite number of local invariant counterterms, including those with the same structure as \eqref{Mop}. In this sense, we can say that \eqref{Mop} is generated at the quantum level. The reason this happens is because $c^{\mu\nu}$, $d^{\mu\nu}$, $e^{\mu}$, $f^\mu$, and $g^{\mu\nu\lambda}$ are dimensionless tensors. The important fact is that all these counterterms involving LSV tensors can be reabsorbed by a non-linear renormalization of the same LSV tensors, see \cite{Sorella,Piguet:1995er}.

\section*{Acknowledgments}
%

%\ni The authors express their gratitude to P. Gaete and A. Spallicci for stimulating discussions on diverse aspects of non-linear electrodynamic models.
%F. Karbstein is also deeply acknowledged for pointing out important references and drawing our attention to a relevant aspect of the Euler-
%Heisenberg effective model. 
%

M. J. Neves thanks CNPq (Conselho Nacional de Desenvolvimento Cient\' ifico e Tecnol\'ogico), Brazilian scientific support federal agency, for partial financial support, Grant number 305743/2026-0 (GM).

%
%\section{References}
%

\end{document}